\documentclass[journal,onecolumn]{IEEEtran}

\usepackage{graphicx}
\usepackage{float}
\usepackage{array}
\usepackage{tabularx}
\usepackage{times}
\usepackage{amsmath}
\usepackage{amssymb}
\usepackage{adjustbox}

\usepackage{verbatim}
\usepackage{moreverb}
\usepackage{alltt}
\usepackage{cite}
\usepackage{epsfig}
\usepackage{subfigure}
\usepackage{multirow}
\usepackage{slashbox}
\usepackage{indentfirst}
\usepackage{epstopdf}
\usepackage{amsfonts}
\usepackage[ruled,vlined]{algorithm2e}
\usepackage{url}
\begin{document}

\title{Exploiting the Unexploited of Coded Caching for Wireless Content Distribution:\\Detailed Theoretical Proofs}

\vspace{-0.25cm}

%\author{\IEEEauthorblockN{Sinong Wang\IEEEauthorrefmark{1}, Xiaohua Tian\IEEEauthorrefmark{1} and Hui Liu\IEEEauthorrefmark{1}}

\author{\IEEEauthorblockN{Sinong Wang, Xiaohua Tian and Hui Liu}

\IEEEauthorblockA{\IEEEauthorrefmark{0}Department of Electronic Engineering, Shanghai Jiao Tong University\\ \{snwang, xtian, huiliu\}@sjtu.edu.cn}
}
\maketitle

{\emph{\textbf{Lemma 1:}}} Under the request vector $\left(d_{1}, d_{2}, \ldots, d_{K}\right)$ and caching distribution $Q$, the traffic rate produced by our encoding and decoding scheme $\Gamma$ is
\begin{equation}
\sum\limits_{i=1}^{K}\sum\limits_{v\subset[K],|v|=i}\max\limits_{j\in v}\{(q_{d_{j}}M)^{i-1}(1-q_{d_{j}}M)^{K-i+1}\}.
\end{equation}
\par{\emph{\textbf{Proof:}}} Consider a particular bit in one of the content, termed as content $i$. Since the prefetching is uniform , by symmetry this bit has probability
\begin{equation*}
p=\frac{C_{q_{i}MF}^{1}}{C_{F}{1}}=q_{i}M,
\end{equation*}
of being prefetched in the cache of any fixed user. Consider now a fixed subset of t out of $K$ users. The probability that this bit is prefetched at exactly those $t$ users is
\begin{equation*}
(q_{i}M)^t(1-q_{i}M)^{K-t}.
\end{equation*}
Hence, the average number of bits of content $i$ that are cached at exactly those $t$ users is
\begin{equation*}
F(q_{i}M)^t(1-q_{i}M)^{K-t}.
\end{equation*}
Since $|U/\{k\}|=k-1$ , the expected size of $V_{k,U/\{k\}}$ is
\begin{equation*}
F(q_{i}M)^{s-1}(1-q_{i}M)^{K-k+1}.
\end{equation*}
Note that, for $F$ large enough, the actual realization of random number of bits in $V_{k,S/\{k\}}$ is in the interval
\begin{equation*}
F(q_{i}M)^{k-1}(1-q_{i}M)^{K-k+1}\pm o(F),
\end{equation*}
with high probability. For simplicity, the $o(F)$ term is ignored in the following derivation.
\par Consider a fixed value of $s$ in Line 7 and a fixed subset $S$ of cardinality $s$. In line 8, server sends
\begin{equation*}
\max\limits_{k \in U}V_{k,U/\{k\}}=F\max\limits_{j\in U}\{(q_{d_{j}}M)^{i-1}(1-q_{d_{j}}M)^{K-i+1}\}.
\end{equation*}
Traversing all subsets $U$ of $[K]$, the total traffic rate produced by Algorithm 1 is
\begin{equation*}
F\sum\limits_{i=1}^{K}\sum\limits_{v\subset[K],|v|=i}\max\limits_{j\in v}\{(q_{d_{j}}M)^{i-1}(1-q_{d_{j}}M)^{K-i+1}\}.
\end{equation*}
$\blacksquare$
\par{\emph{\textbf{Lemma 2:}}} Any request vector $\left(d_{1}, d_{2}, \ldots, d_{K}\right)$ satisfying $\vec{s}=\left(\alpha_{1}, \alpha_{2}, \ldots, \alpha_{N}\right)$ produces the same traffic rate (5) under the encoding and decoding scheme $\Gamma$.
\par{\emph{\textbf{Proof:}}} We consider two requested vectors: $$(d_{1}, \ldots, d_{m}, \ldots, d_{n}, \ldots, d_{K})\quad \text{and}\quad (d_{1}, , \ldots, d_{n}^{*}=d_{m}, \ldots, d_{m}^{*}=d_{n}, \ldots, d_{K}).$$
Remark that these two requested vectors satisfying request situation: $\vec{s}=(\alpha_{1}, \alpha_{2}, \ldots, \alpha_{N})$ and the difference between them is that user $m$ and user $n$ exchanges their requests. Factually, the different request vectors satisfying the same request situation can be converted to each other via finite exchange. Then, we will show that, under these two requested vectors, the traffic rate is equal.
\par The traffic rate under $(d_{1}, \ldots, d_{m}, \ldots, d_{n}, \ldots, d_{K})$ is
\begin{equation*}
R_{s}(\vec{s},Q,\Gamma)=\sum\limits_{i=1}^{K}{R_{m}(\vec{s},Q,\Gamma)+R_{n}(\vec{s},Q,\Gamma)+R_{m,n}(\vec{s},Q,\Gamma)+R_{\o}(\vec{s},Q,\Gamma)}.
\end{equation*}
Where
\begin{align*}
&R_{m}(\vec{s},Q,\Gamma)=\sum\limits_{v\subset[K],|v|=i, n \in v, m \notin v}\max\{(q_{d_{m}}M)^{i-1}(1-q_{d_{m}}M)^{K-i+1}, \max\limits_{j\in v /m}\{(q_{d_{j}}M)^{i-1}(1-q_{d_{j}}M)^{K-i+1}\}\},\\
&R_{n}(\vec{s},Q,\Gamma)=\sum\limits_{v\subset[K],|v|=i, n \notin v, m \in v}\max\{(q_{d_{n}}M)^{i-1}(1-q_{d_{n}}M)^{K-i+1}, \max\limits_{j\in v /n}\{(q_{d_{j}}M)^{i-1}(1-q_{d_{j}}M)^{K-i+1}\}\},\\
&R_{m,n}(\vec{s},Q,\Gamma)=\sum\limits_{v\subset[K],|v|=i, n \notin v, m \in v}\max\left\{ {\begin{array}{*{20}{c}}
(q_{d_{m}}M)^{i-1}(1-q_{d_{m}}M)^{K-i+1},(q_{d_{n}}M)^{i-1}(1-q_{d_{n}}M)^{K-i+1},\\
\max\limits_{j\in v /n}\{(q_{d_{j}}M)^{i-1}(1-q_{d_{j}}M)^{K-i+1}\}
\end{array}} \right\},\\
&R_{\o}(\vec{s},Q,\Gamma)=\sum\limits_{v\subset[K],|v|=i} \max\limits_{j\in v /\{m,n\}}\{(q_{d_{j}}M)^{i-1}(1-q_{d_{j}}M)^{K-i+1}\}.
\end{align*}
\par The traffic rate under $(d_{1}, \ldots, d_{m}^{*}, \ldots, d_{n}^{*}, \ldots, d_{K})$ is
\begin{equation*}
R_{s}^{*}(\vec{s},Q,\Gamma)=\sum\limits_{i=1}^{K}{R_{m}^{*}(\vec{s},Q,\Gamma)+R_{n}^{*}(\vec{s},Q,\Gamma)+R_{m,n}^{*}(\vec{s},Q,\Gamma)+R_{\o}^{*}(\vec{s},Q,\Gamma)}.
\end{equation*}
Where
\begin{align*}
&R_{m}^{*}(\vec{s},Q,\Gamma)=\sum\limits_{v\subset[K],|v|=i, n \in v, m \notin v}\max\{(q_{d_{m}}M)^{i-1}(1-q_{d_{m}}M)^{K-i+1}, \max\limits_{j\in v /m}\{(q_{d_{j}}M)^{i-1}(1-q_{d_{j}}M)^{K-i+1}\}\},\\
&R_{n}^{*}(\vec{s},Q,\Gamma)=\sum\limits_{v\subset[K],|v|=i, n \notin v, m \in v}\max\{(q_{d_{n}}M)^{i-1}(1-q_{d_{n}}M)^{K-i+1}, \max\limits_{j\in v /n}\{(q_{d_{j}}M)^{i-1}(1-q_{d_{j}}M)^{K-i+1}\}\},\\
&R_{m,n}^{*}(\vec{s},Q,\Gamma)=\sum\limits_{v\subset[K],|v|=i, n \notin v, m \in v}\max\left\{ {\begin{array}{*{20}{c}}
(q_{d_{m}}M)^{i-1}(1-q_{d_{m}}M)^{K-i+1},(q_{d_{n}}M)^{i-1}(1-q_{d_{n}}M)^{K-i+1},\\
\max\limits_{j\in v /n}\{(q_{d_{j}}M)^{i-1}(1-q_{d_{j}}M)^{K-i+1}\}
\end{array}} \right\},\\
&R_{\o}^{*}(\vec{s},Q,\Gamma)=\sum\limits_{v\subset[K],|v|=i} \max\limits_{j\in v /\{m,n\}}\{(q_{d_{j}}M)^{i-1}(1-q_{d_{j}}M)^{K-i+1}\}.
\end{align*}
Consider $d_{m}^{*}=d_{n}$ and $d_{n}^{*}=d_{m}$, we can get
\begin{align*}
&R_{m}(\vec{s},Q,\Gamma)=R_{n}^{*}(\vec{s},Q,\Gamma), R_{n}(\vec{s},Q,\Gamma)=R_{m}^{*}(\vec{s},Q,\Gamma),\\
&R_{m,n}^{*}(\vec{s},Q,\Gamma)=R_{n,m}(\vec{s},Q,\Gamma)=R_{m,n}^{*}(\vec{s},Q,\Gamma), R_{\o}(\vec{s},Q,\Gamma)=R_{\o}^{*}(\vec{s},Q,\Gamma).
\end{align*}
Hence,
\begin{equation*}
R_{s}(\vec{s},Q,\Gamma)=R_{s}^{*}(\vec{s},Q,\Gamma).
\end{equation*}
$\blacksquare$

\par{\emph{\textbf{Theorem 3:}}} For $N \in \mathbb{N}$ contents and $K \in \mathbb{K}$ users each with cache size $0 \leq M \leq N$. If $p_{1} \leq p_{2} \leq \cdots \leq p_{N}$ and $q_{1} \leq q_{2} \leq \cdots \leq q_{N}$, then
\begin{equation*}
R^{ub}(P,Q,\Gamma)=F\sum\limits_{i=1}^{N}\mathbb{P}(A_{i})\frac{1-q_{i}M}{q_{i}M}\left[{1-\left( {1-q_{i}M} \right)^{k}}\right].
\end{equation*}
equal if only if $p_{1} = p_{2} = \cdots = p_{N}$ and $q_{1} = q_{2} = \cdots = q_{N}$. Where $\mathbb{P}(A_{i})$ represents the probability that $K$ users request content $i, i+1, \ldots, N$, and can be calculated as
\begin{align*}
&\mathbb{P}(A_{i})=({1-\sum\limits_{j=1}^{i-1}p_{j}})^{K}\cdot [{1-({1-\frac{p_{i}}{1-\sum_{j=1}^{i-1}p_{j}}})^{K}}].
\end{align*}
\par{\emph{Proof:}} All request situation is divided into following $N$ cases:
\begin{align*}
&A_{1}: \alpha_{1}>0, \alpha_{2}>0, \ldots, \alpha_{N}>0;\\
&A_{2}: \alpha_{1}=0, \alpha_{2}>0, \ldots, \alpha_{N}>0;\\
&A_{3}: \alpha_{1}=0, \alpha_{2}=0, \ldots, \alpha_{N}>0;\\
&\qquad\qquad\qquad\cdots\cdots\\
&A_{N}: \alpha_{1}=0, \alpha_{2}=0, \ldots, \alpha_{N}>0.\\
\end{align*}
In the case $A_{i}$, each user only request one of contents $i, i+1, \ldots, N$ and their corresponding caching distribution satisfies $q_{i}\leq q_{i+1} \leq \cdots \leq q_{N}$. Let
\begin{equation*}
q_{i}, q_{i+1}, \cdots, q_{N} \leftarrow q_{i},
\end{equation*}
then the caching distribution of content $i+1, i+2, \ldots, N$ are reduced to $q_{i}$. Thus the traffic rate under this case is
\begin{equation*}
R_{s}(A_{i},Q,\Gamma)=F\sum\limits_{k=1}^{K}C_{K}^{k}(q_{i}M)^{k-1}(1-q_{i}M)^{K-k+1}=F\frac{1-q_{i}M}{q_{i}M}\left(1-(1-q_{i}M)^{K}\right).
\end{equation*}
\par The probability of case $A_{i}$ is calculated based on multiplication formula,
\begin{align*}
P(A_{i})&=P\{\overline{C_{1}} \cdot \overline{C_{2}} \cdots \overline{C_{i}} \cdot C_{i+1}\cdots C_{N}\}\\
&=P\{\overline{C_{1}}\} \cdot P\{\overline{C_{2}}|\overline{C_{1}}\} \cdots P\{\overline{C_{N-1}}|\overline{C_{1}} \cdots \overline{C_{i}} \cdot C_{i+1}\cdots C_{N-2}\} \cdot P\{\overline{C_{N}}|\overline{C_{1}} \cdots \overline{C_{i}} \cdots C_{i+1}\cdots C_{N-1}\}\\
&=\left({1-\sum\limits_{j=1}^{i-1}p_{j}}\right)^{K}\cdot \left[{1-\left({1-p_{i}^{(i-1)}}\right)^{K}}\right].
\end{align*}
Where
\begin{equation*}
p_{i}^{(i-1)}=\frac{p_{i}}{1-\left({1-p_{i}^{(i-1)}}\right)^{K}}.
\end{equation*}
Hence, the upper bound of total traffic rate is
\begin{equation*}
R^{ub}(P,Q)=\sum\limits_{i=1}^{N}P(A_{i})\frac{1-q_{i}M}{q_{i}M}\left[{1-\left( {1-q_{i}M} \right)^{k}}\right].
\end{equation*}
$\blacksquare$

{\emph{\textbf{Theorem 4:} }}For $N \in \mathbb{N}$ contents and $K \in \mathbb{K}$ users each with cache size $0 \leq M \leq N$.
\begin{equation*}
R(P,Q,\Gamma) \geq R^{lb}(P)=\sum\limits_{i=1}^{N}\mathbb{P}(B_{i})\max\limits_{1 \leq k \leq i,K}\left({k-\frac{k}{\lfloor i/k \rfloor}}M\right).
\end{equation*}
where $\mathbb{P}(B_{i})$ denotes the probability that $K$ users only requests $i$ kinds of contents, which can be derived by the concept of generating function.
\par{\emph{\textbf{Proof:}}} Since the size of each content is identical, all request situation is divided into $N$ cases:
\begin{align*}
&B_{1}: \alpha_{i}>0,i=k_{1},\alpha_{i}=0,i\neq k_{1};\\
&B_{2}: \alpha_{i}>0,i=k_{1},k_{2},\alpha_{i}=0,i\neq k_{1},k_{2};\\
&\cdots\cdots\\
&B_{N}: \alpha_{i}>0, i \in \mathbb{K}.
\end{align*}
\par Under the case $i$, there are only $i$ contents are requested by all users. Consider the cut separating $V_{1}, V_{2}, \ldots, V_{\lfloor i/s\rfloor}$ and $Z_{1}, Z_{2}, \ldots, Z_{k}$. By the cut set bound\cite{1},
\begin{equation*}
\lfloor i/s\rfloor R^{lb}(B_{i})+kM \geq k\lfloor i/k\rfloor.
\end{equation*}
Optimizing over all possible choices of $k$, we obtain the lower bound of case $i$.
\begin{equation*}
R^{lb}(B_{i}) \geq \max\limits_{k \in \{1,\ldots,\min\{i,K\}\}}\left(k-\frac{k}{\lfloor i/k \rfloor}\right).
\end{equation*}
Consider all cases, the average lower bound is
\begin{align*}
R^{lb}(P)&=\sum\limits_{i=1}^{N}\mathbb{P}(B_{i})R^{lb}(B_{i})=\sum\limits_{i=1}^{N}\mathbb{P}(B_{i})\max\limits_{k \in \{1,\ldots,\min\{i,K\}\}}\left(k-\frac{k}{\lfloor i/k \rfloor}\right).
\end{align*}
$\blacksquare$

{\emph{\textbf{Lemma 3:}}} When $K=\omega(N^{v})$ and $v>1$ or $K=\Theta(N^{v})$ and $v>1$, the traffic produced by encoding and decoding scheme $\Gamma$ and approximate caching distribution $Q^{\dag}$  satisfies
\begin{equation*}
\lim\limits_{K, N \rightarrow \infty}\frac{R^{ub}(P,Q^{\dag},\Gamma)}{R^{lb}(P)} \leq \frac{M-c}{cM-c^2-c^2M} \cdot \frac{1}{1-e^{\frac{1}{-\zeta(v)}}}.
\end{equation*}
\par{\emph{\textbf{Proof:}}} Consider the first case: $K=\omega(N^{v})$ and $v>1$. Then, the upper bound is
\begin{align}
{R^{ub}}\left( {P,{Q^\dag }, \Gamma} \right) &= \sum\limits_{i = 1}^N {P\left( {{A_i}} \right)} \frac{{1 - q_i^\dag M}}{{q_i^\dag M}}\left[ {1 - {{\left( {1 - q_i^\dag M} \right)}^K}} \right]\notag\\
&= \sum\limits_{i = 1}^N {P\left( {{A_i}} \right)} \left[ {\frac{{\frac{{\left( {N - M} \right){{\left( {\frac{1}{{{p_i}}}} \right)}^{\frac{1}{{K - 1}}}}}}{{\sum\limits_{i = 1}^N {{{\left( {\frac{1}{{{p_i}}}} \right)}^{\frac{1}{{K - 1}}}}} }}}}{{1 - \frac{{\left( {N - M} \right){{\left( {\frac{1}{{{p_i}}}} \right)}^{\frac{1}{{K - 1}}}}}}{{\sum\limits_{i = 1}^N {{{\left( {\frac{1}{{{p_i}}}} \right)}^{\frac{1}{{K - 1}}}}} }}}}} \right] \cdot \left[ {1 - {{\left( {N - M} \right)}^K}\frac{{{{\left( {\frac{1}{{{p_i}}}} \right)}^{\frac{K}{{K - 1}}}}}}{{{{\left[ {\sum\limits_{i = 1}^N {{{\left( {\frac{1}{{{p_i}}}} \right)}^{\frac{1}{{K - 1}}}}} } \right]}^K}}}} \right].
\end{align}
1) Note that $K=\omega(N^v)$, there exists constant $a$, such that $K \geq aN^{v}$.
\begin{align*}
\lim\limits_{N\rightarrow\infty}\left(\frac{1}{p_{i}}\right)^{\frac{1}{K-1}}&=\left(i^v\sum\limits_{j=1}^{N}\frac{1}{j^v}\right)^{\frac{1}{K-1}}=\lim\limits_{N\rightarrow \infty}i^{\frac{v}{K-1}}\cdot \zeta(v)^{\frac{1}{K-1}}.
\end{align*}
Since the $\zeta(v)$ is the Riemann function\cite{2} and satisfies
\begin{equation*}
\lim\limits_{N \rightarrow \infty}\zeta(v)^{\frac{1}{K-1}}=1, v>1.
\end{equation*}
Then,
\begin{align*}
&\lim\limits_{N\rightarrow \infty}i^{\frac{v}{K-1}}\cdot \zeta(v)^{\frac{1}{K-1}} \leq \lim\limits_{N\rightarrow \infty}i^{\frac{v}{aN^v-1}} \leq \lim\limits_{N\rightarrow \infty}N^{\frac{v}{aN^v-1}}=e^{\lim\limits_{N\rightarrow \infty}\frac{vlnN}{aN^v-1}}=1,\\
&\lim\limits_{N\rightarrow \infty}i^{\frac{v}{K-1}}\cdot \zeta(v)^{\frac{1}{K-1}} = \lim\limits_{N\rightarrow \infty}i^{\frac{v}{K-1}} \geq \lim\limits_{N\rightarrow \infty} 1^{\frac{v}{K-1}}=1.
\end{align*}
Hence,
\begin{equation}
\lim\limits_{N\rightarrow\infty}\left(\frac{1}{p_{i}}\right)^{\frac{1}{K-1}}=1.
\end{equation}
2) Note that $K=\omega(N^v)$, there exists constant $a$, such that $K \geq aN^{v}$.
\begin{align*}
&\lim\limits_{N\rightarrow\infty}\sum\limits_{i=1}^{N}\left(\frac{1}{p_{i}}\right)^{\frac{1}{K-1}}=\lim\limits_{N\rightarrow\infty}\left(\sum\limits_{i=1}^{N}i^{\frac{v}{K-1}}\right) \leq \lim\limits_{N\rightarrow\infty}\left(\sum\limits_{i=1}^{N}i^{\frac{v}{aN^{v}-1}}\right) \leq \lim\limits_{N\rightarrow\infty}\left(\sum\limits_{i=1}^{N}N^{\frac{v}{aN^{v}-1}}\right)=\lim\limits_{N\rightarrow\infty}N\cdot N^{\frac{v}{aN^{v}-1}}=N,\\
&\lim\limits_{N\rightarrow\infty}\sum\limits_{i=1}^{N}\left(\frac{1}{p_{i}}\right)^{\frac{1}{K-1}}=\lim\limits_{N\rightarrow\infty}\left(\sum\limits_{i=1}^{N}i^{\frac{v}{K-1}}\right) \geq \lim\limits_{N\rightarrow\infty}\left(\sum\limits_{i=1}^{N}1^{\frac{v}{K-1}}\right)=N.
\end{align*}
Hence,
\begin{equation}
\lim\limits_{N\rightarrow\infty}\sum\limits_{i=1}^{N}\left(\frac{1}{p_{i}}\right)^{\frac{1}{K-1}}=N, 1\leq i \leq N.
\end{equation}
3) Note that $K=\omega(N^v)$, there exists constant $a$, such that $K \geq aN^{v}$.
\begin{align*}
\lim\limits_{N\rightarrow\infty}\left[\sum\limits_{i=1}^{N}\left(\frac{1}{p_{i}}\right)^{\frac{1}{K-1}}\right]^{K}&=\lim\limits_{N\rightarrow\infty}\left(\sum\limits_{i=1}^{N}i^{\frac{v}{K-1}}\right)^{K}\cdot\left(\sum\limits_{i=1}^{N}\frac{1}{i^{v}}\right)^{\frac{K}{K-1}} \\ &=\lim\limits_{N\rightarrow\infty}\left(\sum\limits_{i=1}^{N}i^{\frac{v}{K-1}}\right)^{K}\cdot\zeta(v) \leq \lim\limits_{N\rightarrow\infty}\left(\sum\limits_{i=1}^{N}N^{\frac{v}{K-1}}\right)^{K}\cdot\zeta(v)=\left[\zeta(v)N^{v}\right]\cdot N^{K}.
\end{align*}

Based on H$\ddot{o}$lder inequality\cite{3}
\begin{equation*}
\left( {\sum\limits_{i = 1}^N {a_i^p} } \right) \cdot \left( {\sum\limits_{i = 1}^N {b_i^q} } \right) \ge \sum\limits_{i = 1}^N {{a_i}{b_i}},
\end{equation*}

where $p$ and $q$ satisfies $1/p+1/q=1$. Let $a_{i}=i, b_{i}=1, p=\frac{v}{K-1}$ and $q=1-\frac{K-1}{v}$, we can get
\begin{equation*}
{\left( {\sum\limits_{i = 1}^N {{i^{\frac{v}{{K - 1}}}}} } \right)^{\frac{{K - 1}}{v}}} \cdot {\left( {\sum\limits_{i = 1}^N {{1^{\frac{v}{{v - K + 1}}}}} } \right)^{1 - \frac{{K - 1}}{v}}} \ge \sum\limits_{i = 1}^N i  = \frac{1}{2}N(N + 1).
\end{equation*}

Then,

\begin{align*}
\lim\limits_{N\rightarrow\infty}\left[\sum\limits_{i=1}^{N}\left(\frac{1}{p_{i}}\right)^{\frac{1}{K-1}}\right]^{K}&=\lim\limits_{N\rightarrow\infty}\left(\sum\limits_{i=1}^{N}i^{\frac{v}{K-1}}\right)^{K}\cdot\zeta(v) \geq \left[\zeta(v)\left(\frac{N+1}{2}\right)\right]\cdot N^{K}.
\end{align*}

Hence, there exists $\lambda$ that
\begin{equation}
\lim\limits_{N\rightarrow\infty}\left[\sum\limits_{i=1}^{N}\left(\frac{1}{p_{i}}\right)^{\frac{1}{K-1}}\right]^{K}=\left[\lambda\zeta(v)N^v\right]\cdot N^{K}.
\end{equation}

Based on equations (9)-(12), we can get
\begin{equation}
\lim\limits_{N\rightarrow\infty}{R^{ub}}\left( {P,{Q^\dag }, \Gamma} \right) = \lim\limits_{N\rightarrow\infty} \frac{N-M}{M}\left[1-\frac{1}{\lambda\zeta(v)N^{v}}\left(1-\frac{M}{N}\right)^K\sum\limits_{i=1}^{N}\frac{P(A_{i})}{p_{i}}\right].
\end{equation}

Consider the lower bound
\begin{equation*}
R^{lb}(P)=\sum\limits_{i=1}^{N}\mathbb{P}(B_{i})\max\limits_{1 \leq k \leq \min{i,K}}\left(k-\frac{k}{\lfloor i/k \rfloor}M\right) \geq \sum\limits_{i=1}^{N}\mathbb{P}(B_{i})\max\limits_{1 \leq k \leq \min{i,k}}\left(k-\frac{k^2}{1-k/i}\cdot \frac{M}{i}\right).
\end{equation*}

Let $s=c\frac{i}{M}, 1\leq i \leq N$, where $c$ is a constant between $0$ and $1$, then
\begin{equation*}
R^{lb}(P) \geq \frac{1}{M} \cdot \frac{cM-c^2-c^2M}{M-c} \cdot \sum\limits_{i=1}^{N}i \cdot \mathbb{P}(B_{i}).
\end{equation*}

Remark that $\sum\limits_{i=1}^{N}i \cdot \mathbb{P}(B_{i})$ is the expectation for the number of different contents that $K$ users  request. To avoid the complicated derivation of $\mathbb{P}(B_{i})$, we adopt the indicator function to calculate $\sum\limits_{i=1}^{N}i \cdot \mathbb{P}(B_{i})$.

$X_{i}=\{0,1\}$ represents that content $i$ is requested by at least one user in $K$ users' requests, with probability
\begin{equation*}
\mathbb{P}(X_{i})=1-(1-p_{i})^K.
\end{equation*}

Then, we can get
\begin{equation*}
\sum\limits_{i=1}^{N}i \cdot \mathbb{P}(B_{i})=\sum\limits_{i=1}^{N}\mathbb{E}[X_{i}]=N-\sum\limits_{i=1}^{N}(1-p_{i})^K.
\end{equation*}

Note that
\begin{equation*}
\lim\limits_{N\rightarrow\infty} \sum\limits_{i=1}^{N}(1-p_{i})^K=\lim\limits_{N\rightarrow\infty}\sum\limits_{i=1}^{N}\left(1-\frac{1}{i^v\zeta(v)}\right)^K \leq \lim\limits_{N\rightarrow\infty}\sum\limits_{i=1}^{N}\left(1-\frac{1}{N^v\zeta(v)}\right)^K=\lim\limits_{N\rightarrow\infty}\sum\limits_{i=1}^{N}e^{-\frac{K}{N^v\zeta(v)}} \leq e^{-\frac{a}{\zeta(v)}}N.
\end{equation*}

Thus,
\begin{equation}
\lim\limits_{N\rightarrow\infty}R^{lb}(P) \geq \lim\limits_{N\rightarrow\infty} \frac{N}{M} \cdot \frac{cM-c^2-c^2M}{M-c} \cdot \left(1-e^{-\frac{a}{\zeta(v)}}\right).
\end{equation}

Based on equations (13) and (14), we can get
\begin{align}
\lim\limits_{N\rightarrow\infty}\frac{R^{ub}(P,Q^{\dag},\Gamma)}{R^{lb}(P)} \leq & \lim\limits_{N\rightarrow\infty} \frac{\frac{N-M}{M}\left[1-\frac{1}{\lambda\zeta(v)N^v}\left(1-\frac{M}{N}\right)^{K}\sum\limits_{i=1}^{N}\frac{\mathbb{P}(A_{i})}{p_{i}}\right]}{\frac{cM-c^2-c^2M}{M-c}\cdot\frac{N}{M}\cdot\left(1-e^{-\frac{a}{\zeta(v)}}\right)}\notag\\
=&\lim\limits_{N\rightarrow\infty}\frac{cM-c^2-c^2M}{M-c}\cdot\frac{1}{1-e^{-\frac{a}{\zeta(v)}}}\cdot\left[1-\frac{1}{\lambda\zeta(v)N^v}\left(1-\frac{M}{N}\right)^K\sum\limits_{i=1}^{N}\frac{\mathbb{P}(A_{i})}{p_{i}}\right]\notag\\
\leq & \frac{cM-c^2-c^2M}{M-c}\cdot \frac{1}{1-e^{-\frac{a}{\zeta(v)}}}.
\end{align}

Consider the second case: $K=\Theta(N^v)$ and $v>1$.

According a same procedure in first case, we can get
\begin{align}
&\lim\limits_{N\rightarrow\infty}{R^{ub}}\left( {P,{Q^\dag }, \Gamma} \right) = \lim\limits_{N\rightarrow\infty} \frac{N-M}{M}\left[1-\frac{1}{\lambda\zeta(v)N^{v}}\left(1-\frac{M}{N}\right)^K\sum\limits_{i=1}^{N}\frac{P(A_{i})}{p_{i}}\right]\\
&\lim\limits_{N\rightarrow\infty}R^{lb}(P) \geq \lim\limits_{N\rightarrow\infty} \frac{N}{M} \cdot \frac{cM-c^2-c^2M}{M-c} \cdot \left(1-e^{-\frac{a}{\zeta(v)}}\right).
\end{align}

Hence,
\begin{equation}
\lim\limits_{N\rightarrow\infty}\frac{R^{ub}(P,Q^{\dag},\Gamma)}{R^{lb}(P)} \leq \frac{cM-c^2-c^2M}{M-c}\cdot \frac{1}{1-e^{-\frac{a}{\zeta(v)}}}.
\end{equation}
Let constant $a=1$, we can get the results.\\
$\blacksquare$

{\emph{\textbf{Lemma 4:}}} When $K=\omega(N^v)$ and $v<1$ or $K=\Theta(N^{v})$ and $v<1$, the traffic produced by encoding and decoding scheme $\Gamma$ and approximate caching distribution $Q^{\dag}$  satisfies
\begin{equation*}
\lim\limits_{K, N \rightarrow \infty}\frac{R^{ub}(P,Q^{\dag},\Gamma)}{R^{lb}(P)} \leq \frac{M-c}{cM-c^2-c^2M} \cdot \frac{1}{1-e^{v-1}}.
\end{equation*}
\par{\emph{\textbf{Proof:}}} Consider the first case: $K=\omega(N^{v})$, $v<1$ and a same procedure in Lemma 3. Then, we calculate the upper bound.

1) Note that $K=\omega(N^v)$, there exists constant $a$, such that $K \geq aN^{v}$.
\begin{align*}
&\lim\limits_{N\rightarrow\infty}\left(\frac{1}{p_{i}}\right)^{\frac{1}{K-1}}=\left(i^v\sum\limits_{j=1}^{N}\frac{1}{j^v}\right)^{\frac{1}{K-1}}=\lim\limits_{N\rightarrow \infty}i^{\frac{v}{K-1}}\cdot \sum\limits_{j=1}^{N}\frac{1}{j^v}^{\frac{1}{K-1}} \leq \lim\limits_{N\rightarrow\infty}N^{\frac{v}{K-1}}\cdot N^{\frac{1}{K-1}} \leq \lim\limits_{N\rightarrow\infty} N^{\frac{v+1}{N^V-1}}=1,\\
&\lim\limits_{N\rightarrow\infty}\left(\frac{1}{p_{i}}\right)^{\frac{1}{K-1}} = \lim\limits_{N\rightarrow \infty}i^{\frac{v}{K-1}}\cdot \sum\limits_{j=1}^{N}\frac{1}{j^v}^{\frac{1}{K-1}} \geq  \lim\limits_{N\rightarrow \infty} 1^{\frac{v}{K-1}} \cdot N^{\frac{1-v}{K-1}} = 1.
\end{align*}

Hence,
\begin{equation}
\lim\limits_{N\rightarrow\infty}\left(\frac{1}{p_{i}}\right)^{\frac{1}{K-1}}=1, 1 \leq i \leq N.
\end{equation}

2) Note that $K=\omega(N^v)$, there exists constant $a$, such that $K \geq aN^{v}$.
\begin{align*}
&\lim\limits_{N\rightarrow\infty}\sum\limits_{i=1}^{N}\left(\frac{1}{p_{i}}\right)^{\frac{1}{K-1}}=\lim\limits_{N\rightarrow\infty}\left(\sum\limits_{i=1}^{N}i^{\frac{v}{K-1}}\right) \cdot \left(\sum\limits_{i=1}^{N}\frac{1}{i^v}\right) \leq \lim\limits_{N\rightarrow\infty}\left(\sum\limits_{i=1}^{N}N^{\frac{v}{K-1}}\right) \cdot \left(\sum\limits_{i=1}^{N}\frac{1}{1^v}\right) = \lim\limits_{N\rightarrow\infty}N\cdot N^{\frac{v}{K-1}} \cdot N^{\frac{1}{K-1}} \leq N,\\
&\lim\limits_{N\rightarrow\infty}\sum\limits_{i=1}^{N}\left(\frac{1}{p_{i}}\right)^{\frac{1}{K-1}}=\lim\limits_{N\rightarrow\infty}\left(\sum\limits_{i=1}^{N}i^{\frac{v}{K-1}}\right) \cdot \left(\sum\limits_{i=1}^{N}\frac{1}{i^v}\right) \geq \lim\limits_{N\rightarrow\infty}\left(\sum\limits_{i=1}^{N}1^{\frac{v}{K-1}}\right) \cdot \left(\sum\limits_{i=1}^{N}\frac{1}{N^v}\right)=N.
\end{align*}

Hence,
\begin{equation}
\lim\limits_{N\rightarrow\infty}\sum\limits_{i=1}^{N}\left(\frac{1}{p_{i}}\right)^{\frac{1}{K-1}}=N, 1\leq i \leq N.
\end{equation}

3) It is same as (12), we can also find that there exists constant $\lambda$,
\begin{equation}
\lim\limits_{N\rightarrow\infty}\left[\sum\limits_{i=1}^{N}\left(\frac{1}{p_{i}}\right)^{\frac{1}{K-1}}\right]^{K}=\left[\lambda\sum\limits_{i=1}^{N}i^{-v}N^v\right]\cdot N^{K}.
\end{equation}

Based on equations (19)-(21), we can get
\begin{equation}
\lim\limits_{N\rightarrow\infty}{R^{ub}}\left( {P,{Q^\dag }, \Gamma} \right) = \lim\limits_{N\rightarrow\infty} \frac{N-M}{M}\left[1-\frac{1}{\lambda \sum\limits_{i=1}^{N}i^{-v}N^{v}}\left(1-\frac{M}{N}\right)^K\sum\limits_{i=1}^{N}\frac{P(A_{i})}{p_{i}}\right].
\end{equation}

Consider the lower bound
\begin{equation*}
\lim\limits_{N\rightarrow\infty}R^{lb}(P) \geq \lim\limits_{N\rightarrow\infty} \frac{N}{M} \cdot \frac{cM-c^2-c^2M}{M-c}.
\end{equation*}

Let $k=c\frac{i}{M}, 1\leq i \leq N$, where $c$ is a constant between $0$ and $1$, then
\begin{equation*}
R^{lb}(P) \geq \frac{1}{M} \cdot \frac{cM-c^2-c^2M}{M-c} \cdot \sum\limits_{i=1}^{N}i \cdot \mathbb{P}(B_{i}) \cdot \left[1-\frac{\sum\limits_{i=1}^{N}(1-p_{i})^{K}}{N}\right].
\end{equation*}

Note that
\begin{align*}
\lim\limits_{N\rightarrow\infty} \sum\limits_{i=1}^{N}(1-p_{i})^K&=\lim\limits_{N\rightarrow\infty}\sum\limits_{i=1}^{N}\left(1-\frac{1}{i^v\sum\limits_{j=1}{N}j^{-v}}\right)^K \leq \lim\limits_{N\rightarrow\infty}\sum\limits_{i=1}^{N}\left(1-\frac{1}{N^v\sum\limits_{j=1}^{N}j^{-v}}\right)^K \\ &= \lim\limits_{N\rightarrow\infty}\sum\limits_{i=1}^{N}e^{-\frac{K}{N^v\sum\limits_{j=1}^{N}j^{-v}}} \leq e^{v-1}N.
\end{align*}

According to the limits $\lim\limits_{N\rightarrow\infty}\sum\limits_{i=1}^{N}i^{-v}=\frac{1}{1-v}N^{1-v}$, then
\begin{equation*}
\lim\limits_{N\rightarrow\infty} \sum\limits_{i=1}^{N}(1-p_{i})^K \leq e^{v-1}N.
\end{equation*}

Thus,
\begin{equation}
\lim\limits_{N\rightarrow\infty}R^{lb}(P) \geq \lim\limits_{N\rightarrow\infty} \frac{N}{M} \cdot \frac{cM-c^2-c^2M}{M-c} \cdot \left(1-e^{v-1}\right)/
\end{equation}

Based on equations (22) and (23), we can get
\begin{align}
\lim\limits_{N\rightarrow\infty}\frac{R^{ub}(P,Q^{\dag},\Gamma)}{R^{lb}(P)} \leq \frac{cM-c^2-c^2M}{M-c}\cdot \frac{1}{1-e^{v-1}}.
\end{align}

Consider the second case: $K=\Theta(N^v)$ and $v<1$.

According a same procedure in first case, we can get
\begin{align}
&\lim\limits_{N\rightarrow\infty}{R^{ub}}\left( {P,{Q^\dag }, \Gamma} \right) = \lim\limits_{N\rightarrow\infty} \frac{N-M}{M}\left[1-\frac{1}{\lambda \sum\limits_{i=1}^{N}i^{-v}N^{v}}\left(1-\frac{M}{N}\right)^K\sum\limits_{i=1}^{N}\frac{P(A_{i})}{p_{i}}\right],\\
&\lim\limits_{N\rightarrow\infty}R^{lb}(P) \geq \lim\limits_{N\rightarrow\infty} \frac{N}{M} \cdot \frac{cM-c^2-c^2M}{M-c} \cdot \left(1-e^{v-1}\right).
\end{align}

Hence,
\begin{equation}
\lim\limits_{N\rightarrow\infty}\frac{R^{ub}(P,Q^{\dag},\Gamma)}{R^{lb}(P)} \leq \frac{cM-c^2-c^2M}{M-c}\cdot \frac{1}{1-e^{v-1}}.
\end{equation}
$\blacksquare$

{\emph{\textbf{Lemma 5:}}} When $K=O(N^{v})$ and $v>1$ or $v<1$, the traffic produced by encoding and decoding scheme $\Gamma$ and approximate caching distribution $Q^{\dag}$  satisfies
\begin{equation*}
\lim\limits_{K, N \rightarrow \infty}\frac{R^{ub}(P,Q^{\dag},\Gamma)}{R^{lb}(P)} \leq \frac{M-c}{cM-c^2-c^2M} \cdot \frac{1}{1-\lambda'}.
\end{equation*}
where $c$ is a constant between 0 and 1, $\zeta(v)$ denotes the riemann function.
\par{\emph{\textbf{Proof:}}} The procedure is same as Lemma 3 and Lemma 4.

 \end{document}